\documentclass{physeauth}
\usepackage{graphicx}
\usepackage{amsmath}
\usepackage{amssymb}
\usepackage{psfrag}
\usepackage{epsfig}

\newcommand{\nn}{\nonumber}
\newcommand{\bra}{\langle}
\newcommand{\ve}{\vert}
\newcommand{\ket}{\rangle}

\begin{document} 
\begin{frontmatter}

\title{Nano-Thermodynamics: On the minimal length scale for the existence of temperature}\author[dlr,uni]{Michael Hartmann \thanksref{thank1}},
\author[uni]{G\"unter Mahler} and
\author[ati]{Ortwin Hess}
\address[dlr]{Institut f\"ur Technische Physik, DLR-Stuttgart,
Pfaffenwaldring 38-40, 70569 Stuttgart, Germany}
\address[uni]{Institut f\"ur Theoretische Physik I, Universit\"at Stuttgart,
Pfaffenwaldring 57, 70550 Stuttgart, Germany}
\address[ati]{Advanced Technology Institute, University of Surrey, Guildford GU2 7XH, United Kingdom}
\thanks[thank1]{email: {\tt michael.hartmann@dlr.de}}

\begin{abstract}
We consider a quantum system consisting of a regular chain of elementary subsystems with nearest
neighbor interactions and assume that the total system is in a canonical state with temperature $T$.
We analyze under what condition the state factors into a product of canonical density matrices
with respect to groups of $n$ subsystems each, and when these groups have the same temperature $T$.
While in classical mechanics the validity of this procedure only depends on the size of the groups $n$,
in quantum mechanics the minimum group size $n_{\text{min}}$ also depends on the temperature $T \,$!
As examples, we apply our analysis to a harmonic chain and an Ising spin chain.
We discuss various features that show up due to the characteristics of the models considered.
For the harmonic chain, which successfully describes thermal properties of insulating solids,
our approach gives a first quantitative estimate of the minimal length scale on which temperature can exist:
This length scale is found to be constant for temperatures above the Debye temperature
and proportional to $T^{-3}$ below.
\end{abstract}

\end{frontmatter}


\section{Introduction}

The microscopic limit of the applicability of Thermodynamics is still poorly understood
\cite{GemmerOtte2001,Allahverdyan2000}.
Down to which length scales can its standard concepts meaningfully be defined and employed?
However, despite its fundamental relevance this topic has only received little attention for a long
time \cite{Hill1994,Rajagopal2004}.

The situation drastically changed in recent years, since the issue became highly relevant for
experiments.
Significant progress in the synthesis and processing of materials with structures on
nanometer length scales has created a demand for better understanding of thermal properties of
nanoscale devices, individual nanostructures and nanostructured materials
\cite{Cahill2003}.
Experimental techniques have improved to such an extent that the measurement of thermodynamic
quantities like temperature with a spatial resolution on the nanometer scale seems within reach
\cite{Schwab2000,Gao2002,Pothier1997,Aumentado2001}.

To provide a basis for the interpretation of present day and future experiments in nanoscale physics
and technology and to obtain a better understanding of the limits of thermodynamics,
it is thus indispensable to clarify the applicability of thermodynamical concepts
on small length scales starting from the most fundamental theory at hand, i. e. quantum mechanics.

Here we focus on the question whether temperature can be meaningfully defined on those small length scales.

The existence of thermodynamical quantities \cite{Nieuwenhuizen1998},
i. e. the existence of the thermodynamic limit strongly
depends on the correlations between the considered parts of a system.
Hence, local temperature is only expected to exist in systems with short range interactions, where
the interaction between a region in space and its surrounding only takes place at the surface of that region.
Being only a surface effect, correlations between the region and the surrounding thus become less important
as the size of the region increases, since the energy contained in the region grows with its volume
(diameter$^3$) while correlations only grow with the surface (diameter$^2$)
\cite{Ruelle1969}.

This scaling of interactions between parts of a system \cite{Hartmann2003a}
thus sets a minimal length scale on which correlations are
still small enough to permit the definition of local temperatures.
In this paper we study this connection quantitatively.

We adopt here the convention that a local temperature exists if the considered part of the system is in a
canonical state, where the distribution is an exponentially decaying function of energy characterized by one
single parameter. This implies that there is a one-to-one mapping between temperature and
the expectation values of observables, by which temperature is usually measured. Temperature measurements
based on different observables will thus yield the same result, contrary to distributions with several
parameters \cite{Hartmann2004c}. In large systems composed of very many subsystems, the density of states is a strongly growing
function of energy \cite{Tolman1967}. If the distribution were not exponentially decaying,
the product of the density of states times the distribution would not have a pronounced peak
and thus physical quantities like energy would not have ``sharp'' values.

Based on the above arguments and noting that a quantum description becomes imperative at nanoscopic scales,
the following approach appears to be reasonable:
Consider a large homogeneous quantum system, brought into a thermal state via interaction with its environment,
divide this system into subgroups and analyze for what
subgroup-size the concept of temperature is still applicable.

Harmonic lattice models are a standard tool for the description of thermal properties of solids.
We therefore apply our theory to a harmonic chain model to get estimates that are expected to be relevant
for real materials and might be tested by experiments.

Recently, spin chains have been subject of extensive studies in condensed matter physics and
quantum information theory.
Thus correlations and possible local temperatures in spin chains are of interest, both
from a theoretical and experimental point of view \cite{Osterloh2002,Wang2002,Kenzelmann2002}.
We study spin chains with respect to our present purpose and compare their characteristics with the
harmonic chain.

%
%
\section{General Theory\label{general}} 

We consider a homogeneous (i.e. translation invariant) chain of elementary quantum
subsystems with nearest neighbor interactions.
The Hamiltonian of our system is thus of the form
\begin{equation}\label{hamil}
H = \sum_{i} H_i + I_{i,i+1} \, ,
\end{equation}
where the index $i$ labels the elementary subsystems. $H_i$ is the Hamiltonian of subsystem $i$
and $I_{i,i+1}$ the interaction between subsystem $i$ and $i+1$.
We assume periodic boundary conditions.

We now form $N_G$ groups of $n$ subsystems each
(index $i \rightarrow (\mu-1) n + j; \: \mu = 1, \dots, N_G; \: j = 1, \dots, n$)
and split this Hamiltonian into two parts,
\begin{equation}
\label{hsep}
H = H_0 + I,
\end{equation}
where $H_0$ is the sum of the Hamiltonians of the isolated groups,
\begin{eqnarray}\label{isogroups}
H_0 & = & \sum_{\mu=1}^{N_G} \left( \mathcal{H}_{\mu} - I_{\mu n,\mu n + 1} \right) \enspace \enspace
\textrm{with} \nn \\
\mathcal{H}_{\mu} & = & \sum_{j=1}^n H_{n (\mu - 1) + j} + I_{n (\mu-1) + j,\, n (\mu-1) + j + 1} 
\end{eqnarray}
and $I$ contains the interaction terms of each group with its neighbor group,
\begin{equation}
I = \sum_{\mu=1}^{N_G} I_{\mu n,\mu n + 1}.
\end{equation}
The eigenstates of the Hamiltonian $H_0$ are products of group eigenstates,
\begin{equation}
\label{prodstate}
H_0 \: \ve a \ket = E_a \: \ve a \ket \quad \text{with} \quad
\ve a \ket = \prod_{\mu = 1}^{N_G} \ve a_{\mu} \ket,
\end{equation}
where
$\left( \mathcal{H}_{\mu} - I_{\mu n, \mu n + 1} \right) \ve a_{\mu} \ket = E_{\mu} \ve a_{\mu} \ket$.
$E_{\mu}$ is the energy of one subgroup only and $E_a = \sum_{\mu=1}^{N_G} E_{\mu}$.

\subsection{Thermal State in the Product Basis}
We assume that the total system is in a thermal state
\begin{equation}
\label{candens}
\hat \rho = \frac{e^{- \beta H}}{Z},
\end{equation}
where $Z$ is the partition sum and $\beta = (k_B T)^{-1}$
the inverse temperature with Boltzmann's constant $k_B$ and temperature $T$.

We are interested in the matrix representation of the state (\ref{candens}) in the product basis,
the eigenbasis of $H_0$. The diagonal elements are the expectation values of the density operator
(\ref{candens}) in the states $\ve a \ket$,
\begin{equation}
\label{newrho}
\bra a \ve \hat \rho \ve a \ket =
\int_{E_0}^{E_1} w_a (E) \: \frac{e^{- \beta E}}{Z} \: dE,
\end{equation}
where $E_0$ is the energy of the ground state and $E_1$ the upper limit of the spectrum \cite{IntLim}.
$w_a (E)$ is defined as follows: The state $\ve a \ket$ is not an eigenstate of the total Hamiltonian $H$.
Thus, if $H$ would be measured in the state $\ve a \ket$, eigenvalues of $H$ would be obtained with
certain probabilities: $w_a (E)$ is the density of this probability distribution.
Since the Hamiltonian $H$ is the sum of Hamiltonians of the groups, the situation has some
analogies to a sum of random variables. This indicates that there might exist a central limit theorem
for the present quantum system, provided the number of groups becomes very large \cite{Billingsley1995} .
Since the state $\ve a \ket$ is not translation invariant and since $H$ also contains the group
interactions, the central limit theorem has to be of a Lyapunov (or Lindeberg)
type for mixing sequences \cite{Linnik1971}.
One can indeed show that such a quantum central limit theorem exists for the present model
\cite{Hartmann2003,Hartmann2004b} and that $w_a (E)$ thus converges to a Gaussian normal distribution
in the limit of infinite number of groups $N_G$,
\begin{equation}
\label{gaussian_dist}
\lim_{N_G \to \infty} w_a (E) = \frac{1}{\sqrt{2 \pi} \Delta_a}
\exp \left(- \frac{\left(E - E_a - \varepsilon_a \right)^2}{2 \, \Delta_a^2} \right),
\end{equation}
where the quantities $\varepsilon_a$ and $\Delta_a$ are defined by 
\begin{eqnarray}
\varepsilon_a & \equiv & \bra a \ve H \ve a \ket - E_a \\
\Delta_a^2 & \equiv & \bra a \ve H^2 \ve a \ket - \bra a \ve H \ve a \ket^2.
\end{eqnarray}
$\varepsilon_a$ is the difference between the energy expectation value of the distribution $w_a (E)$ and the
energy $E_a$, while $\Delta_a^2$ is the variance of the energy $E$ for the distribution $w_a (E)$.
Note that $\varepsilon_a$ has a classical counterpart while $\Delta_a^2$ is purely quantum mechanical.
It appears because the commutator $[H,H_0]$ is nonzero, and the distribution $w_a(E)$ therefore has nonzero
width. The two quantities $\varepsilon_a$ and $\Delta_a^2$ can also be expressed in terms of the
interaction only (see eq. (\ref{hsep})),
\begin{eqnarray}
\varepsilon_a & = & \bra a \ve I \ve a \ket\\
\Delta_a^2 & = & \bra a \ve I^2 \ve a \ket - \bra a \ve I \ve a \ket^2,
\end{eqnarray}
meaning that $\varepsilon_a$ is the expectation value and $\Delta_a^2$ the squared width of the interactions
in the state $\ve a \ket$.

The rigorous proof of equation (\ref{gaussian_dist}) is given in \cite{Hartmann2003} and based on
the following two assumptions:
The energy of each group $\mathcal{H}_{\mu}$ as defined in equation (\ref{isogroups}) is bounded, i. e.
\begin{equation}
\label{bounded}
\bra \chi \ve \mathcal{H}_{\mu} \ve \chi \ket \le C
\end{equation}
for all normalized states $\ve \chi \ket$ and some constant $C$, and
\begin{equation}
\label{vacuumfluc}
\bra a \ve H^2 \ve a \ket - \bra a \ve H \ve a \ket^2 \ge N_G \, C'
\end{equation}
for some constant $C' > 0$.

In scenarios where the energy spectrum of each elementary subsystem has an upper limit, such as spins,
condition (\ref{bounded}) is met a priori.

For subsystems with an infinite energy spectrum, such as harmonic oscillators,
we restrict our analysis to states where the energy of every group,
including the interactions with its neighbors, is bounded. Thus, our considerations do not apply
to product states $\ve a \ket$, for which all the energy was located in only one group or only a small
number of groups. Since $N_G  \gg 1$, the number of such states is vanishingly small compared
to the number of all product states.

If conditions (\ref{bounded}) and (\ref{vacuumfluc}) are met, equation (\ref{newrho})
can be computed for $N_G  \gg 1$ \cite{Hartmann2004}:
\begin{equation}
\label{newrho2}
\begin{split}
\bra a \ve \hat \rho \ve a \ket = \frac{1}{2 \, Z} \,
\exp \left(- \beta \overline{E}_a + \frac{\beta^2 \Delta_a^2}{2} \right) \times \hspace{2cm} \\[0.2cm]
\left[\textrm{erfc} \left( \frac{E_0 - \overline{E}_a + \beta \Delta_a^2}{\sqrt{2} \, \Delta_a} \right) -
\textrm{erfc} \left( \frac{E_1 - \overline{E}_a + \beta \Delta_a^2}{\sqrt{2} \, \Delta_a} \right) \right]
\end{split}
\end{equation}
where $\overline{E}_a \equiv \bra a \ve H \ve a \ket = E_{a} + \varepsilon_a$ and $\textrm{erfc} (x)$ is the
conjugate Gaussian error function \cite{Abramowitz1970}.
The second error function in (\ref{newrho2}) only appears if the energy is bounded and the
integration extends from the energy of the ground state $E_0$ to the upper limit of the spectrum $E_1$.

Note that $\overline{E}_a$ is a sum of $N_G$ terms and that $\Delta_a$ fulfills equation (\ref{vacuumfluc}).
The arguments of the conjugate error functions thus grow proportional to $\sqrt{N_G}$ or stronger.
If these arguments divided by $\sqrt{N_G}$ are finite (different from zero),
the asymptotic expansion of the error function \cite{Abramowitz1970} may thus be used for $N_G \gg 1$:
$\text{erfc}(x) \approx \exp \left(- x^2 \right) / \left(\sqrt{\pi} \, x \right)$ for $x \rightarrow \infty$
and $\text{erfc}(x) \approx 2 + \exp \left(- x^2 \right) / \left(\sqrt{\pi} \, x \right)$ for
$x \rightarrow - \infty$.
Inserting this approximation into equation (\ref{newrho2}) and using $E_0 < \overline{E}_a < E_1$ shows
that the second conjugate error function, which contains the upper limit of the energy spectrum,
can always be neglected compared to the first, which contains the ground state energy.

The off diagonal elements $\bra a \ve \hat \rho \ve b \ket$ vanish for\linebreak
$\ve E_a - E_b \ve > \Delta_a + \Delta_b$ because the overlap of the two distributions of conditional
probabilities becomes negligible. For $\ve E_a - E_b \ve < \Delta_a + \Delta_b$, the transformation
involves an integral over frequencies and thus these terms are significantly smaller than
the entries on the diagonal.

\subsection{Conditions for Local Thermal States}
We now test under what conditions the density matrix $\hat \rho$ may be approximated by a product
of canonical density matrices with temperature $\beta_{\text{loc}}$ for each subgroup $\mu = 1, 2, \dots, N_G$.
Since the trace of a matrix is invariant under basis transformations, it is sufficient to verify
the correct energy dependence of the product density matrix.
If we assume periodic boundary conditions,
all reduced density matrices are equal and their product
is of the form $\bra a \ve \hat \rho \ve a \ket \propto \exp(- \beta_{\text{loc}} E_a)$.
We thus have to verify whether the logarithm of rhs of equation (\ref{newrho2})
is a linear function of the energy $E_a$, 
\begin{equation} \label{log}
\ln \left( \bra a \ve \hat \rho \ve a \ket \right) \approx - \beta_{\text{loc}} \, E_a + c,
\end{equation}
where $\beta_{\text{loc}}$ and $c$ are constants.

Applying the asymptotic expansion of the error function in  equation (\ref{newrho2}),
one realizes that equation (\ref{log}) can only be true for
\begin{equation} \label{cond_const}
\frac{E_a + \varepsilon_a  - E_0}{\sqrt{N_G} \, \Delta_a} > \beta \frac{\Delta_a^2}{\sqrt{N_G} \, \Delta_a} ,
\end{equation}
and
\begin{equation}
\label{cond_linear_1} 
- \varepsilon_a + \frac{\beta}{2} \, \Delta_a^2 \approx  c_1 E_a + c_2 ,
\end{equation}
where $c_1$ and $c_2$ are arbitrary real constants.
Note that $\varepsilon_a$ and $\Delta_a^2$ need not be functions of $E_a$ and therefore in general
cannot be expanded in a Taylor series.

To ensure that the density matrix of each subgroup $\mu$ is approximately canonical, one needs to satisfy
(\ref{cond_linear_1}) for each subgroup $\mu$ separately;
\begin{equation}
\label{cond_linear_2} 
- \frac{\varepsilon_{\mu - 1} + \varepsilon_{\mu}}{2} + \frac{\beta}{4} \,
\left(\Delta_{\mu - 1}^2 + \Delta_{\mu}^2 \right)
+ \frac{\beta}{6} \, \tilde{\Delta}_{\mu}^2 \, \approx \, c_1  \, E_{\mu} + c_2
\end{equation}
where $\varepsilon_{\mu} = \bra a \ve I_{\mu n, \mu n + 1} \ve a \ket$,
$\Delta_{\mu}^2 = \bra a \ve \mathcal{H}_{\mu}^2 \ve a \ket - \bra a \ve \mathcal{H}_{\mu} \ve a \ket^2$\linebreak and
$\tilde{\Delta}_{\mu}^2 = \sum_{\nu = \mu-1}^{\mu+1} \, \bra a \ve \mathcal{H}_{\nu-1} \mathcal{H}_{\nu} +
\mathcal{H}_{\nu} \mathcal{H}_{\nu-1} \ve a \ket -$\linebreak
$-2 \, \sum_{\nu = \mu-1}^{\mu+1}
\bra a \ve \mathcal{H}_{\nu-1} \ve a \ket \bra a \ve \mathcal{H}_{\nu} \ve a \ket$.

Temperature becomes intensive, if the constant $c_1$ vanishes,
\begin{equation} \label{intensivity}
\left| c_1 \right| \ll 1 \enspace \enspace \Rightarrow \enspace \enspace \beta_{\text{loc}} = \beta.
\end{equation}
If this was not the case, temperature would not be intensive, although it might exist locally.

It is sufficient to satisfy conditions (\ref{cond_const}) and (\ref{cond_linear_2}) for an adequate energy
range $E_{\text{min}} \le E_{\mu} \le E_{\text{max}}$ only.
The density of states of large many body systems is
typically a rapidly growing function of energy \cite{GemmerPhD,Tolman1967}. If the total system is in a
thermal state, occupation probabilities decay exponentially with energy. The product of these two
functions is thus sharply peaked at the expectation value of the energy $\overline{E}$ of the total
system $\overline{E} + E_0 = $Tr$(H \hat \rho)$, with $E_0$ being the ground state energy.
The energy range needs to be centered around this peak and large enough.
On the other hand it must not be larger than the range of values $E_{\mu}$ can take on.
Therefore a pertinent and ``safe'' choice for $E_{\text{min}}$ and $E_{\text{max}}$ is
\begin{equation} \label{e_range}
\begin{array}{rcl}
E_{\text{min}} & = & \text{max}
\left( \left[E_{\mu}\right]_{\text{min}} \, , \,
\frac{1}{\alpha} \frac{\overline{E}}{N_G} + \frac{E_0}{N_G} \right)\\[0.4cm]
E_{\text{max}} & = & \text{min}
\left( \left[E_{\mu}\right]_{\text{max}} \, , \, \alpha
\frac{\overline{E}}{N_G} + \frac{E_0}{N_G} \right) \, ,
\end{array}
\end{equation}
where $\alpha \gg 1$ and $\overline{E}$ will in general depend on the global temperature.
In equation (\ref{e_range}), $\left[E_{\mu}\right]_{\text{min}}$ and $\left[E_{\mu}\right]_{\text{max}}$ denote
the minimal and maximal values $E_{\mu}$ can take on.

For a model obeying equations (\ref{bounded}) and (\ref{vacuumfluc}), the two conditions
(\ref{cond_const}) and (\ref{cond_linear_2}), which constitute the general result of this article,
must both be satisfied. These fundamental criteria will now be applied to some concrete examples.
%
%

\section{Results for Special Models}
\label{sec:results-spec-models}
We are now going to examine equations (\ref{cond_const}) and (\ref{cond_linear_2}) for some special
models.
As discussed in the introduction, those terms in (\ref{cond_const}) and (\ref{cond_linear_2}),
which contain the interactions between neighboring groups ($\varepsilon_{\mu}$ and $\Delta_{\mu}$)
are independent of the group size, while the ``background'' term $E_{\mu}$ grows approximately
linear with the group size $n$. Hence, the interaction terms become less important compared to $E_{\mu}$
with increasing group size. Therefore, taking into account the energy range (\ref{e_range}),
conditions (\ref{cond_const}) and (\ref{cond_linear_2}) can be used to determine a minimal
group size $n_{\text{min}}$. Technical details of this calculation can be found
in \cite{Hartmann2004} and \cite{Hartmann2004a}.

For each model, the results depend, besides the global temperature $\beta$, on two ``accuracy'' parameters,
$\alpha$ and $\delta$, which quantify the tolerated deviations from a canonical distribution.
The first parameter, $\alpha$, has already been introduced in equation (\ref{e_range}).
It is a measure for the energy range, where one demands (\ref{cond_const}) and (\ref{cond_linear_2})
to be fulfilled.
The second parameter, $\delta$, is the ratio between the terms in the lhs of condition (\ref{cond_linear_2}),
which are linear in $E_{\mu}$,  and those, which are of higher order. Thus $\delta = 0.01$ means that
all terms in the lhs of (\ref{cond_linear_2}) which are not linear in $E_{\mu}$ are at least two orders of
magnitude smaller than the linear ones.

\subsection{Harmonic Chain\label{harmonicchain}} 

As a representative for the class of systems with an infinite energy spectrum,
we consider a harmonic chain of $N_G \cdot n$ particles of mass $m$ and
spring constant $\sqrt{m} \, \omega_0$. In this case, the Hamiltonian reads
\begin{eqnarray}
H_i & = & \frac{m}{2} \, p_i^2 + \frac{m}{2} \, \omega_0^2 \, q_{i}^2 \\
I_{i, i+1} & = & - m \, \omega_0^2 \, q_{i} \, q_{i+1},
\end{eqnarray}
where $p_i$ is the momentum of the particle at site $i$ and $q_{i}$ the displacement from its equilibrium
position $i \cdot a_0$ with $a_0$ being the distance between neighboring particles at equilibrium.
We divide the chain into $N_G$ groups of $n$ particles each and thus get
a partition of the type considered above. Describing the groups in Debye approximation (see \cite{Hartmann2004}
for details) we calculate the minimal group size $n_{\text{min}}$.
 
\begin{figure}[h]
\psfrag{-4.1}{\small \raisebox{-0.1cm}{$10^{-4}$}}
\psfrag{-3.1}{\small \raisebox{-0.1cm}{$10^{-3}$}}
\psfrag{-2.1}{\small \raisebox{-0.1cm}{$10^{-2}$}}
\psfrag{-1.1}{\small \raisebox{-0.1cm}{$10^{-1}$}}
\psfrag{1.1}{\small \raisebox{-0.1cm}{$10^{1}$}}
\psfrag{2.1}{\small \raisebox{-0.1cm}{$10^{2}$}}
\psfrag{3.1}{\small \raisebox{-0.1cm}{$10^{3}$}}
\psfrag{4.1}{\small \raisebox{-0.1cm}{$10^{4}$}}
\psfrag{1}{}
\psfrag{2}{\small \hspace{+0.2cm} $10^{2}$}
\psfrag{3}{}
\psfrag{4}{\small \hspace{+0.2cm} $10^{4}$}
\psfrag{5}{}
\psfrag{6}{\small \hspace{+0.2cm} $10^{6}$}
\psfrag{7}{}
\psfrag{8}{\small \hspace{+0.2cm} $10^{8}$}
\psfrag{n}{\raisebox{0.1cm}{$n_{\text{min}}$}}
\psfrag{c1}{$\: T / \Theta$}
\epsfig{file=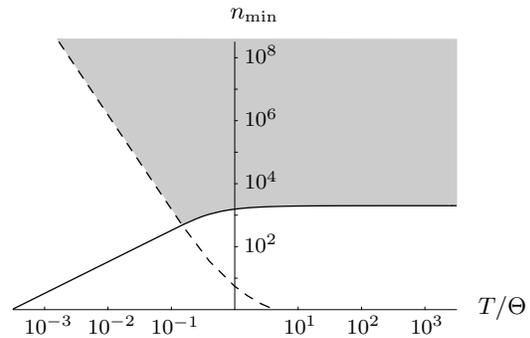,width=7cm}
\caption{Log-log-plot of $n_{\text{min}}$ from eq. (\ref{cond_const}) (dashed line) and
$n_{\text{min}}$ from eq. (\ref{cond_linear_2}) (solid line) for $\alpha = 10$ and $\delta = 0.01$
as a function of $T / \Theta$ for a harmonic chain. $\Theta$ is the Debye temperature,
a characteristic constant of the respective material, which can be found tabulated \cite{Kittel1983}. $\delta$ and $\alpha$ are explained
at the beginning of section \ref{sec:results-spec-models}.
Local temperature exists in the shaded area.}
\label{temp}
\end{figure}

Here, we only present the results for the present model, which are shown in figure \ref{temp}.

%
%
%

For high (low) temperatures $n_{\text{min}}$ can be estimated by
\begin{equation} \label{eq:1}
n_{\text{min}} \approx \left\{
\begin{array}{lcr}
2 \, \alpha / \delta & \textrm{for} & T > \Theta\\
\left( 3 \alpha / 2 \pi^2 \right) \, \left( \Theta / T \right)^3 & \textrm{for} & T < \Theta
\end{array}
\right. \, ,
\end{equation}
where $\Theta$ is the Debye temperature,
a characteristic constant of the respective material, which can be found tabulated \cite{Kittel1983}.

As an extra result, we get, that whenever local temperature exists, it is equal to the global one, i. e. temperature is intensive.
%
%
\subsection{Ising Spin Chain in a Transverse Field\label{isingchain}} 

In this section we consider an Ising spin chain in a transverse field. For this model the Hamiltonian
reads,
\begin{eqnarray} \label{ising_ham}
H_i & = & - B \, \sigma_i^z \nn \\
I_{i, i+1} & = & - \frac{J}{2} \, \left( \sigma_i^x \otimes \sigma_{i+1}^x -
\sigma_i^y \otimes \sigma_{i+1}^y \right) \, ,
\end{eqnarray}
where $\sigma_i^x, \sigma_i^y$ and $\sigma_i^z$ are the Pauli matrices. $B$ is the magnetic field
and $J$ a coupling parameter. We will always assume $B > 0$.

\begin{figure}[h]
\psfrag{-6.1}{\small \raisebox{-0.1cm}{$10^{-6}$}}
\psfrag{-4.1}{\small \raisebox{-0.1cm}{$10^{-4}$}}
\psfrag{-2.1}{\small \raisebox{-0.1cm}{$10^{-2}$}}
\psfrag{2.1}{\small \raisebox{-0.1cm}{$10^{2}$}}
\psfrag{4.1}{\small \raisebox{-0.1cm}{$10^{4}$}}
\psfrag{2}{\small \hspace{+0.2cm} $10^{2}$}
\psfrag{4}{\small \hspace{+0.2cm} $10^{4}$}
\psfrag{6}{\small \hspace{+0.2cm} $10^{6}$}
\psfrag{8}{\small \hspace{+0.2cm} $10^{8}$}
\psfrag{10}{\small \hspace{+0.35cm} $10^{10}$}
\psfrag{12}{\small \hspace{+0.35cm} $10^{12}$}
\psfrag{n}{\raisebox{0.1cm}{$n_{\text{min}}$}}
\psfrag{c1}{$\: T / B$}
\epsfig{file=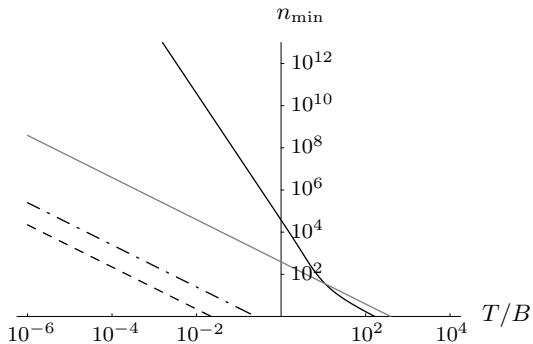,width=7cm}
\caption{Log-log-plot of $n_{\text{min}}$ for $J = 0.1 \times B$ from eq. (\ref{cond_const}) (dashed line),
and from eq. (\ref{cond_linear_2}) (dash - dotted line) and $n_{\text{min}}$ for $J = 10 \times B$
from eq. (\ref{cond_const}) (solid line) and from eq. (\ref{cond_linear_2}) (gray line)
as a function of $T / B$. $\alpha = 10$ and
$\delta = 0.01$. $\delta$ and $\alpha$ are explained
at the beginning of section \ref{sec:results-spec-models}.}
\label{L=0}
\end{figure}

Running through a similar calculation as for the harmonic chain, the details of which can be found
in \cite{Hartmann2004a}, we get the result shown in figure \ref{L=0}.

%
%
%

As for the harmonic chain, temperature is intensive whenever it exists locally, i.e. then
local temperatures are equal to global ones.

%
\section{Summary and Conclusions\label{conclusion}}

We have considered a linear chain of particles interacting with their nearest neighbors.
We have partitioned the chain into identical groups of $n$ adjoining particles each.
Taking the number of such groups to be very large and assuming the total
system to be in a thermal state with temperature $T$ we have found conditions
(equations (\ref{cond_const}) and (\ref{cond_linear_2})), which ensure that each group is approximately
in a thermal state. Furthermore, we have determined when the isolated groups have the same temperature $T$,
that is when temperature is intensive.

The result shows that, in the quantum regime,
these conditions depend on the temperature $T$, contrary to the classical case.
The characteristics of the temperature dependence are determined by the width $\Delta_a$ of the distribution
of the total energy eigenvalues in a product state, which in turn is related to the fact that
the commutator $[H,H_0]$ is nonzero. In particular, the ground state of the total
system does not factorize with respect to any partition
\cite{Wang2002,Vidal2003,Jordan2003,Allahverdyan2002}.

We have then applied the general method to a harmonic chain and an Ising spin chain in a magnetic field.
When applied to specific models, conditions (\ref{cond_const}) and (\ref{cond_linear_2}) determine a 
minimal group size $n_{\text{min}}$ for which a local canonical state and hence local temperature can exist.
We have given order of magnitude estimates of $n_{\text{min}}$ for these models.
Grains of a smaller size are no more in a thermal state and temperature measurements
with a higher resolution should therefore no longer be interpreted in a standard way.

For the spin chain and the harmonic chain, the temperature dependencies of $n_{\text{min}}$
for low temperatures coincide, $n_{\text{min}} \propto T^{-3}$, because both couplings have the
same structure and the upper limit of the spectrum of the spin chain becomes irrelevant at
low temperatures.   

On the other hand, the high temperature asymptotics of $n_{\text{min}}$ differs significantly:

For spins at very high global temperatures, the total density matrix is almost completely mixed, i. e.
proportional to the identity matrix, and thus does not change under basis transformations.
There are thus global temperatures which are high enough, so that local temperatures exist even for
single spins.

For the harmonic chain, this feature does not appear, since the size of the relevant energy range
increases indefinitely with growing global temperature, leading to the constant minimal length
scale in the high energy range.

Since harmonic lattice models in Debye approximation have proven to be successful
in modeling thermal properties of insulators (e.g. heat capacity) \cite{Kittel1983},
our calculation for the harmonic chain provides a first estimate of the minimal length scale on which intensive
temperatures exist in insulating solids,
\begin{equation}
\label{length}
l_{\text{min}} = n_{\text{min}} \, a_0.
\end{equation}
Let us give some numerical estimates: Choosing the ``accuracy parameters''
to be $\alpha = 10$ and $\delta = 0.01$,
we get for hot iron ($T \gg \Theta \approx 470 \,$K, $a_0 \approx 2.5 \,${\AA})
$l_{\text{min}} \approx 50 \,\mu$m, while
for carbon ($\Theta \approx 2230 \,$K, $a_0 \approx 1.5 \,${\AA}) at room temperature ($270 \,$K)
$l_{\text{min}} \approx 10 \,\mu$m. The coarse-graining will experimentally be most relevant at very low
temperatures,
where $l_{\text{min}}$ may even become macroscopic. A pertinent example is silicon
($\Theta \approx 645 \,$K, $a_0 \approx 2.4 \,${\AA}),
which has $l_{\text{min}} \approx 10 \,$cm at $T \approx 1 \,$K
(again with $\alpha = 10$ and $\delta = 0.01$).

Of course the validity of the harmonic lattice model
will eventually break down at finite, high temperatures and our estimates will thus no longer apply there.

Measurable consequences of the local breakdown of the concept of temperature are interesting questions which
arise in the context of the present discussion.

In the scenarios of global equilibrium, which we consider here,
a temperature measurement with a microscopic thermometer, locally in thermal contact with
the large chain, would not reveal the non existence of local temperature.
One can model such a measurement with a small system, representing the thermometer,
coupled to a heat bath, representing the chain. It is a known result of such system bath
models \cite{Weiss1999},
that the system always relaxes to a thermal state with the global temperature
of the bath, no matter how local the coupling might be.

This, however, does not mean that the existence or non existence of local temperatures
had no physical relevance: There are indeed physical properties, which are determined
by the local states rather than the global ones. Whether these properties are of
thermal character depends on the existence of local temperatures.
A detailed discussion of such properties will be given elsewhere \cite{Hartmann2004c}. 

The length scales, calculated in this paper, should also constrain the way one can meaningfully
define temperature profiles in non-equilibrium scenarios \cite{Michel2003}. Here,
temperature measurements with a microscopic thermometer, which is locally in thermal
contact with the sample, might indeed be suitable to measure the local temperature.
An explicit study of this possibility should be subject of future research.

We thank M.\ Michel, M.\ Henrich, H.\ Schmidt, M.\ Stollsteimer, F.\ Tonner and C.\ Kostoglou
for fruitful discussions.

%
%


\begin{thebibliography}{99}
\bibitem{IntLim}
For systems with an energy spectrum that does not have an upper bound, the limit
$E_1 \rightarrow \infty$ should be taken.
\end{thebibliography}
\end{document}